\documentclass[aps,epsfig,twocolumn]{revtex4}

\usepackage{graphicx}
\usepackage{dcolumn}
\usepackage{amssymb}
\usepackage{amsmath}
\usepackage{bm}
\usepackage[dvips]{color}

\newcommand{\bq}{\begin{equation}}
\newcommand{\eq}{\end{equation}}
\newcommand{\bqa}{\begin{eqnarray}}
\newcommand{\eqa}{\end{eqnarray}}

\begin{document}

\title{ 
{
Many-body generalization of the $Z_2$ topological invariant for
the quantum spin Hall effect }}

\author{ Sung-Sik Lee$^{1,2}$ and Shinsei Ryu$^1$ }
\affiliation{
$^1$ Kavli Institute for Theoretical Physics,
University of California, Santa
Barbara, California 93106, {USA} \\
$^2$ Department of {Physics and Astronomy}, McMaster University,
Hamilton, Ontario L8S 4M1, Canada
}

\date{\today}

\begin{abstract}
We propose a many-body generalization of the $Z_2$ topological invariant
for the quantum spin Hall insulator, which does not rely on single-particle
band structures. The invariant is derived as a topological obstruction
that distinguishes topologically distinct many-body ground states
on a torus. It is also expressed as a Wilson-loop of the SU(2) Berry gauge
field, which is quantized due to time-reversal symmetry.
\end{abstract}

\maketitle

A topological insulator is a quantum phase of matter with a bulk energy gap
that cannot be deformed continuously to a trivial band
insulator without going through a quantum phase transition
\cite{Wen90}.
The most prominent example is the quantum Hall (QH) state
\cite{Klitzing,Tsui,Laughlin}.
One physical manifestation of topological orders in QH states is the
existence of gapless edge modes, which are robust against any small
perturbations
\cite{Halperin82,Wen92}.

Topological orders in QH states are associated with the charge
degrees of freedom of electrons {and} accompanied by broken
time-reversal symmetry (TRS).
Recently, Kane and Mele
have established a topological classification in
time-reversal (TR) invariant {noninteracting} systems
associated with the spin degrees of freedom
\cite{KaneMele}.
Considering the doubled version of the Haldane model
for the integer QH effect \cite{Haldane88} as an example, they
showed that there exists a new type of topological insulators,
quantum spin Hall (QSH) insulators,
which are characterized by an odd number of
Kramers pairs of gapless edge modes. In contrast, trivial insulators
have an even number of pairs. QSH insulators
in the bulk are characterized by a $Z_2$ number
\cite{KaneMele,Roy06,Moore06,Fu06,Essin07}.

Although the QSH insulator has been proposed in a {noninteracting} system,
the even or odd parity in the number of Kramers pairs of edge modes is robust
against weak interactions that respect TRS
\cite{KaneMele,CongjunWu06,Xu06}.
This suggests that
the $Z_2$ topological order in the bulk is also robust against those
perturbations.
It is the purpose of this paper to construct a many-body generalization of
the $Z_2$ invariant that detects such bulk $Z_2$ topology.
When interactions are weak, the construction we propose coincides with
(and hence is a natural generalization of)
the known $Z_2$ invariant in {noninteracting limits.}
While it is applicable even in the presence of strong interactions,
our scheme does not exclude the possibility
that the construction of the $Z_2$ invariant
might not be unique,
and hence the corresponding strongly interacting many-body state can be
characterized by
{a multiplet} of $Z_2$ invariants.

\begin{figure}[h!]
        \includegraphics[height=3.5cm,width=8cm]{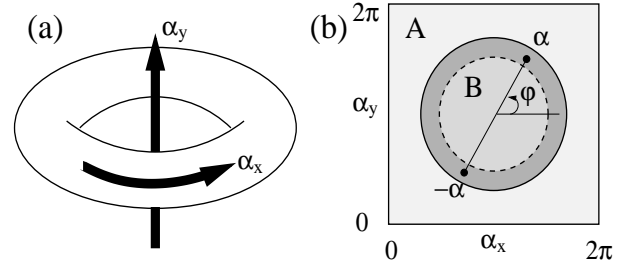} 
\caption{ 
(a) The torus in real space where magnetic fluxes
$\alpha_{x,y}$ are threaded in the $x, y$-cycles, respectively.
(b) The space of the Hamiltonian $T^2_{\alpha}$ obtained by the
flux-threading. The doublet $v^A(\alpha)$ ($v^B(\alpha)$) is well-defined in
the patch A(B).}
\label{fig:1}
\end{figure}

We put a many-body system with a fixed number of particles on a 2D torus $T^2$
with $N = L_x \times L_y$ unit cells, each of which has $m$ sites,
where $N$ is assumed to be odd.
We focus on the case with $m=2$ which includes the Kane-Mele model generalized by adding interactions.
Consider a deformation of the Hamiltonian by threading magnetic fluxes $0\le \alpha_{x,y} < 2\pi$
through the two cycles of the torus where $\alpha_x$ ($\alpha_y$) denotes the flux threaded along
{the $x(y)$ directions} [Fig.\ \ref{fig:1} (a)].
The space of the fluxes forms a torus, which we call
$T^2_\alpha$ to distinguish it from the torus $T^2$
in real space [Fig.\ \ref{fig:1} (b)].
We consider a $2N$-particle ground state.
If there exists a finite energy gap above the ground state
at $\alpha=0$, there is a gap everywhere in $T^2_{\alpha}$
for a large enough system, as the effects of
the fluxes becomes smaller for larger system\cite{CommentSpinFlux}.
Therefore there exists a unique $2N$-particle ground state $|\Psi^\alpha \rangle$
with a finite energy gap at each point $\alpha \in T^2_\alpha$.

The QH states are characterized by a topological obstruction of the U(1) bundle on $T^2_{\alpha}$,
i.e., inability to smoothly define the ground state over $T^2_\alpha$,
which is signaled by a U(1) Chern number
\cite{Thouless82,Niu85}.
On the other hand, in the QSH effect, the U(1) Chern number always vanishes because
of TRS.
However, a non-trivial \textit{substructure} can still be hidden in the
ground state.
For example, when the $z$-component of spin $(S^z)$ is conserved,
the Chern numbers $\mathrm{Ch}_{\uparrow, \downarrow}$
for up and down spins can be defined separately,
which are related by $\mathrm{Ch}_{\uparrow}= -\mathrm{Ch}_{\downarrow}$
because of TRS.
The parity (even or odd) of
$\mathrm{Ch}_{\uparrow}$
defines the $Z_2$ invariant in this limiting situation.

To probe such a non-trivial substructure of $|\Psi^{\alpha}\rangle$,
we consider the following $N$-electron states obtained by creating $N$ holes
in the ground state,
\begin{eqnarray}
|t, s (\alpha) \rangle = \left( \prod\nolimits_{r=1}^N c^{\ }_{r, t, s} \right)
|\Psi^\alpha \rangle,
\label{eq:1}
\end{eqnarray}
where $c^{\dag}_{r,t,s}/c^{\ }_{r,t,s}$ creates/annihilates an electron
with spin $s=\uparrow,\downarrow$ at a site $t=1,2$
within a unit cell labeled by $r$.
If $d(\alpha) := \det \langle 1, s | 1, s' \rangle \neq 0$,
the vector space spanned by the states
$ | 1, \uparrow (\alpha) \rangle$ and
$ | 1, \downarrow (\alpha) \rangle$
is two-dimensional,
and a doublet of orthonormal states
$ v(\alpha) \equiv (| 1 (\alpha) \rangle , |2 (\alpha) \rangle)$
can be obtained from
suitable linear combinations,
$| n (\alpha) \rangle =
\sum_{s=\uparrow,\downarrow} u_{ns} (\alpha) | 1, s (\alpha) \rangle$
with $n=1,2$.
$u_{ns} (\alpha)$ is chosen so that
$ \langle n (\alpha) |m (\alpha) \rangle = \delta_{n,m}$.
From the doublet, we can define a projection operator $\hat P(\alpha) =
|1(\alpha) \rangle \langle 1(\alpha)| + |2(\alpha) \rangle \langle
2(\alpha)|$.

In the case of non-interacting electrons (such as the Kane-Mele model),
the doublet $v(\alpha)$ can be constructed from the Bloch wavefunctions,
and the specific way of creating the doublet, e.g.,
choice of the sublattice we made ($t=1$) at which we remove electrons,
does not matter, i.e., different choices of sublattice lead to
the same projection operator. When we perturb the system by weak interactions,
since $d(\alpha)$ cannot vanish abruptly, the above construction of the doublet $v(\alpha)$
smoothly interpolates non-interacting and interacting cases. However, in general,
$d(\alpha)$ can vanish at some points in $T^2_\alpha$ where the two states $ | 1, s (\alpha) \rangle$
with $s=\uparrow,\downarrow$
fail to span a 2D vector space.
Since $d(\alpha)$ is a real function defined on the 2D space $T^2_\alpha$, $d(\alpha)$ can vanish either
on lines or points in $T^2_\alpha$.
$d(\alpha)$ cannot vanish on a 2D submanifold of $T^2_\alpha$ unless $d(\alpha)$ is
identically zero since the ground state $| \Psi^\alpha \rangle$, and hence $d(\alpha)$ as well, are analytic on $T^2_\alpha$.
If $d(\alpha)$ vanishes at all points in $T^2_\alpha$, we can perturb our Hamiltonian slightly without
closing the gap so that $d(\alpha)$ becomes nonzero everywhere except for some lines and points.
Since the dimension of the submanifold where $d(\alpha)$ vanishes is smaller than $2$, we can
analytically continue the 2D projection operator $\hat P(\alpha)$ to the whole $T^2_\alpha$ generically.
One may think that such a smooth continuation may not work because there can be points
$\alpha_i$ at which one (or a linear combination) of the doublet $|1,s (\alpha)\rangle$ vanishes either due to
a scalar vortex (winding in the phase of the coefficient of a quantum state with a
vanishing norm at the center), i.e., $
| 1, s(\alpha) \rangle =
[ (\alpha_x-\alpha_{ix}) + i ( \alpha_y-\alpha_{iy}) ]
| \psi_{i} \rangle
$
or due to a vector vortex (winding between two quantum states with a vanishing norm at
the center), i.e., $
| 1, s(\alpha) \rangle =
  (\alpha_x-\alpha_{ix}) |\psi_{i1} \rangle
+ (\alpha_y-\alpha_{iy}) |\psi_{i2} \rangle
$ where $|\psi_i\rangle$ and $|\psi_{ia} \rangle$ are well-defined
states with $\langle \psi_{i1} | \psi_{i2} \rangle = 0$. For scalar vortices, $\hat P(\alpha)$ can be smoothly extended because
the projection operator is insensitive to the overall phase. On the other hand,
for vector vortices $\hat P(\alpha)$ cannot be smoothly defined. However, one can generically avoid
the occurrence of the latter points by adding a small TR symmetric perturbation to the
Hamiltonian. If we add a perturbation, the points where the norm vanishes will disappear
because the state will generically change as $|1,s(\alpha) \rangle \rightarrow |1,s(\alpha) \rangle + A(\alpha) |\psi_{i}' \rangle$
where $A(\alpha)\neq 0$ at $\alpha=\alpha_i$ and $| \psi_{i}' \rangle$ is generically independent with $| \psi_{ia} \rangle$ and $|1,-s(\alpha_i)\rangle$.
In this way, the 2D projection operator $\hat P(\alpha)$ can be smoothly defined in the whole $T^2_\alpha$.

We now choose $v(\alpha)$
such that it satisfies the TRS condition,
$v(-\alpha) ( i \sigma_2) = \hat \Theta v(\alpha)$,
where $\hat \Theta$ is the TR operator
\cite{convention}.
Note that when acting on $v(\alpha)$,
which consists of $N$-electron states, $\Theta^2=-1$.
While this is always possible \textit{locally} since
$\hat \Theta \hat P(\alpha) \hat \Theta^{-1} = \hat P(-\alpha)$,
there is no guarantee that
such basis can be defined globally.
Therefore, we first divide $T^2_\alpha$
into two patches ({$A$ and $B$}) as shown in Fig.\ \ref{fig:1} (b),
and see
whether we can merge them into a single patch or not.
In each patch,
we have a doublet
$
v^p(\alpha)=
(
| 1^p (\alpha) \rangle , |2^p (\alpha) \rangle
)
$
($p=A,B$),
satisfying
$\langle n^p(\alpha) | m^p(\alpha) \rangle =
\delta_{nm}$,
$v^p(-\alpha) ( i \sigma_2) = \hat \Theta v^p(\alpha)$,
and
$\hat P (\alpha) =
|1^p(\alpha) \rangle \langle 1^p(\alpha)|
+ |2^p(\alpha) \rangle \langle
2^p(\alpha)|$.

\begin{figure}[h!]
        \includegraphics[height=3.5cm,width=8cm]{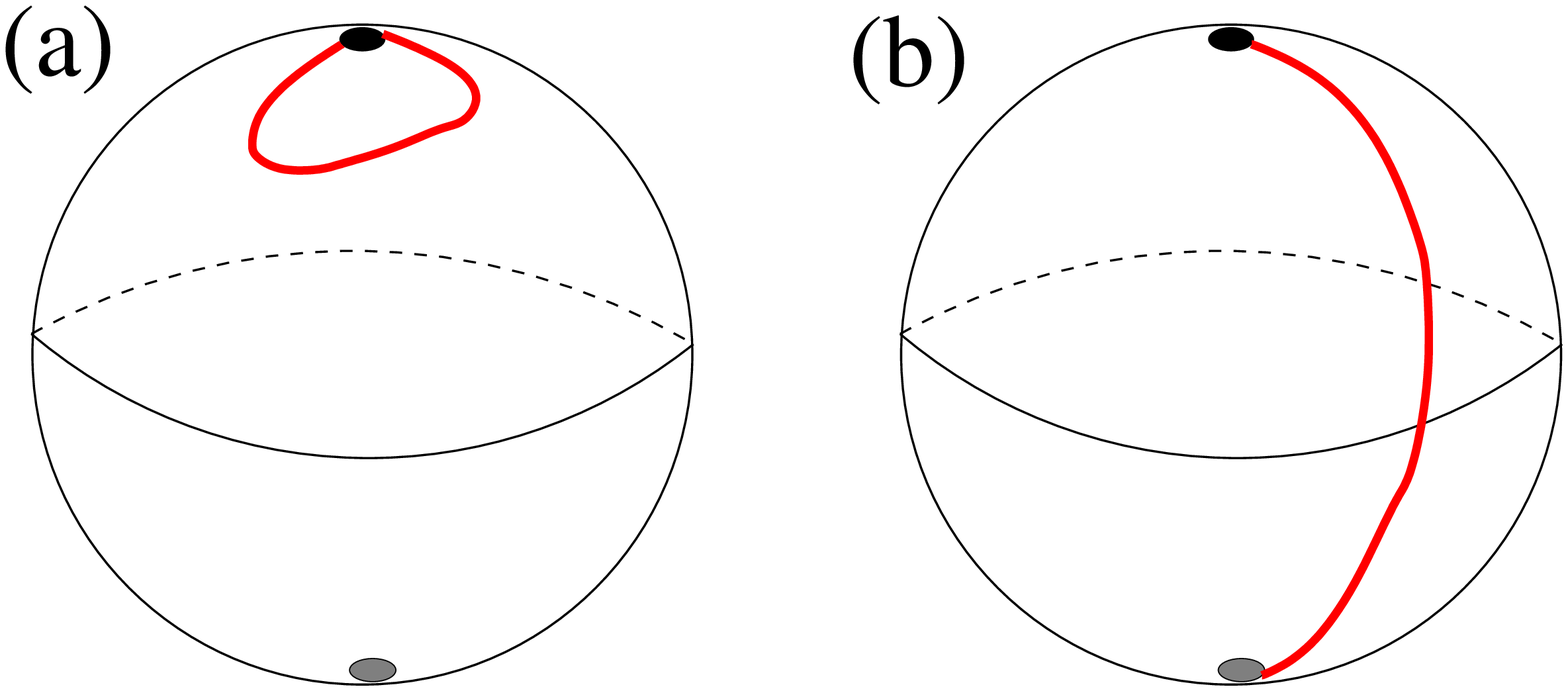} \caption{Two topologically different paths the SU(2) matrix
$g(\alpha)$ can take as the angle $\varphi$ in $A \cap B$ changes from $0$
to $\pi$. Here the ``sphere'' represents the space of SU(2) matrices,
$S^3$. The path in (a) is contractible while the path in (b) is not
because $g(-\alpha)$ should be at the opposite point of the $g(\alpha)$
in $S^3$. }
\label{fig:2}
\end{figure}

Since $v^A(\alpha)$ and $v^B(\alpha)$ span the
same 2D Hilbert space at $\alpha \in A \cap B$, they are
related by
$
v^A (\alpha) = v^B(\alpha)
M(\alpha)
$,
where $M(\alpha)$ is a U(2) matrix
(transition function).
Using the TRS condition,
we obtain a relation
between the transition function at $-\alpha$ and $\alpha$,
$M(-\alpha) = \sigma_2 M^*(\alpha) \sigma_2$.
If we decompose $M(\alpha)$
into a U(1) phase $\theta(\alpha)$
and an SU(2) matrix $g(\alpha)$, we have the two
possibilities,
\begin{eqnarray}
 a) & \theta(-\alpha) = -\theta(\alpha) & ~\mbox{and}
~~ g(-\alpha) = g(\alpha),
\nonumber \\
b) & \theta(-\alpha) = -\theta(\alpha) +
\pi & ~ \mbox{and} ~~ g(-\alpha) = - g(\alpha).
\label{g}
\end{eqnarray}
In the TR invariant system, the U(1) phase cannot have a non-trivial winding.
However, the SU(2) matrix has two topologically distinctive
configurations.
In the case a), we can deform a ground state continuously such that the
trajectory of $g(\alpha)$ as a function of $\varphi$,
where $\varphi$ parameterizes the overlapping region $A \cap B$
which has the topology of $S^1$,
is contracted to a point in the space of SU(2) matrices as shown in
Fig.\ \ref{fig:2} (a).
On the other hand, in the case b), the trajectory
of $g(\alpha)$ is not contractible because $g(\alpha) = - g(-\alpha)$ as
is shown in Fig. \ref{fig:2} (b). Therefore, a state in the class b)
cannot be continuously deformed to a state in class a). The relative
sign between $g(\alpha)$ and $g(-\alpha)$ is the $Z_2$ topological
invariant. If there is no interaction, $\alpha$ plays the role of
momenta of Bloch states and the topological invariant reduces to the
existing $Z_2$ invariant
\cite{Roy06}.
In the present scheme, the $Z_2$
invariant can be generalized to interacting cases. Since the
generalized $Z_2$ invariant is quantized, it cannot change abruptly
upon turning on interactions.

Then when can the $Z_2$ invariant change, as we tune some parameters (other than $\alpha$)
of the Hamiltonian, along a path in the space of Hamiltonians?
The $Z_2$ invariant is well-defined if the following three conditions are satisfied:
(1) there is TRS,
(2) the ground state is uniquely defined, and
(3) $\hat P(\alpha)$ is well-defined at all points in $T^2_\alpha$.
Accordingly, there are three possibilities where the $Z_2$ invariant may change.
The two obvious cases are a breaking of TRS and a phase transition. The last
possibility is the case where $\hat P(\alpha)$ becomes ill-defined without encountering either
of the former two situations either because $d(\alpha)$ identically vanishes or because
there occurs vector vortices. However, if $\hat P(\alpha)$ is not well-defined at a point on
the path in the space of Hamiltonians, one can detour the point by slightly modifying
the path so that $\hat P(\alpha)$ becomes well-defined at all points along a new path. This is
always possible because the measure of the points where $d(\alpha)$ identically vanishes or
vector vortices occur is zero in the space of TR symmetric Hamiltonians.
Namely, the {codimension} of the subspace where $\hat P(\alpha)$ becomes ill-defined is infinite.
This guarantees that the space of well-defined $\hat P(\alpha)$ is connected
as long as there is no phase transition.
Since the
quantized $Z_2$ invariant is well-defined along the new path, the invariant should be
the same for the two end points. Therefore, the $Z_2$ invariant can change only when there
is TRS breaking or a phase transition between the two points. Thus, two quantum
states with different $Z_2$ invariants are always topologically distinct. On the other
hand, if they have the same $Z_2$ invariant, it is generically unclear if they are
adiabatically connected or not. From this point of view, it would be interesting if
there is a many-body state which has a non-trivial $Z_2$ invariant, and yet is separated
from the known non-interacting $Z_2$ insulator by a phase transition.

The non-trivial $Z_2$ topological order amounts to our inability to define the doublet globally while maintaining the TRS condition
$v(-\alpha) ( i \sigma_2 ) = \hat \Theta v(\alpha)$.
If we relax this condition, the doublet can always be
defined in a single patch.
For a trivial case, this is true by definition,
whereas for a non-trivial $Z_2$ insulator, this can be done by defining
$v(\alpha) = v^A(\alpha)$ for $\alpha \in A$ and
$v(\alpha) = v^B(\alpha) \tilde M(\alpha)$
for $\alpha \in B$ where
$\tilde M(\alpha)=M(\alpha)$
for $\alpha \in A \cap B$
and $\tilde M(\alpha)=1$ at $\alpha=0$. A smooth function
$\tilde M(\alpha)$ which satisfies the above conditions can always
be found because $\Pi_1(S^3) = \o$,
where $\Pi_1$ is the first homotopy group
and $S^3$ is the space of the SU(2) group.
The price to pay for making such a single patch
is that we inevitably {lose} the TRS condition
relating $v(\alpha)$ and $v(-\alpha)$,
$v(-\alpha) ( i \sigma_2 ) = \hat \Theta v(\alpha)$,
if the $Z_2$ invariant is non-trivial.
Instead, they are related by
$v( -\alpha) w(\alpha) = \hat \Theta v(\alpha)$,
where $w(\alpha)$ is a general U(2) matrix.
Since $\hat \Theta^2=-1$
$w(\alpha)$ must
satisfy
$w^T(\alpha) = -w(-\alpha)$.
Especially, at the four TR symmetric points, $(0,0)$, $(\pi,0)$,
$(\pi,\pi)$ and $(0,\pi)$, $w(\alpha)$ is antisymmetric. From the two
orthonormal states $| n(\alpha)\rangle$ with $n=1,2$ at each $\alpha$,
we can define a U(2) Berry gauge field,
$A^{mn}_\mu(\alpha) d \alpha^{\mu}=
\langle m(\alpha)| d |n(\alpha)\rangle$.
The U(2) gauge field can uniquely be decomposed
into the U(1) ($a^0_{\mu}$) and SU(2) ($\boldsymbol{a}_{\mu}$) parts as
$A_{\mu} = a^{0}_{\mu}/(2i) + \boldsymbol{a}_{\mu}\cdot
\boldsymbol{\sigma}/(2i) $.

The condition that sews local frames at $\alpha$ and $-\alpha$ together
naturally induces a constraint on the Berry
gauge field configuration,
\begin{eqnarray}
\label{eq:sewing condition for U2 guage field}
A^{\ }_{\mu}(-\alpha) = w(\alpha) A^T_{\mu}(\alpha) w^{\dag}(\alpha) -
w(\alpha)\partial_{\mu}w^{\dag}(\alpha).
\end{eqnarray}
Accordingly, the U(1) and SU(2) gauge fields are constrained as
$
a^{0}_{\mu}(-\alpha) = a^{0}_{\mu}(\alpha) -2
\partial_{\mu}\zeta(\alpha)$,
and
$
\boldsymbol{a}_{\mu}(-\alpha)\cdot \boldsymbol{\sigma}/(2i) =
\boldsymbol{a}_{\mu}(\alpha) \cdot \tilde{w}(\alpha)
\boldsymbol{\sigma}^T \tilde{w}^{\dag}(\alpha) /(2i)
-
\tilde{w}(\alpha)\partial_{\mu} \tilde{w}^{\dag}(\alpha),
$
where we decomposed $w(\alpha)$ into the U(1) ($e^{i\zeta}$) and SU(2) ($\tilde{w}$) parts,
$w(\alpha)=e^{i\zeta(\alpha)} \tilde{w}(\alpha)$.
(Note that this decomposition has a global sign ambiguity, which will
not affect the following discussions.) At the TR symmetric points,
$\tilde{w}(\alpha)$ is equal to $i\sigma_2$ {up to sign},
$
\tilde{w}(\alpha) = \mathrm{Pf}\,[\tilde{w}(\alpha)] \times i \sigma_2
$,
where $\mathrm{Pf}\,[\tilde{w}]$ is the Pfaffian of $\tilde{w}$.

\begin{figure}[h!]
        \includegraphics[height=6cm]{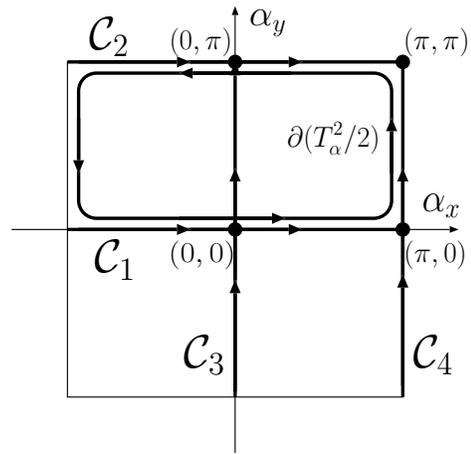} \caption{  Four time-reversal invariant loops $\mathcal{C}_{1,2,3,4}$
on $T^2_{\alpha}$, and a loop $\partial(T^2_{\alpha}/2)$ that encloses
the half of $T^2_{\alpha}$.} \label{fig:3}
\end{figure}

Following Ref.\ \onlinecite{Fu06}, a natural object to consider, which is
invariant under $\alpha$-dependent U(2) transformations of the local
frame, is an SU(2) Wilson loop
\begin{eqnarray}
W[\mathcal{C}] =
\frac{1}{2} \mathrm{Tr}\,P \exp\left[
\oint_{\mathcal{C}} d\alpha^{\mu} \boldsymbol{a}_{\mu}(\alpha) \cdot
\frac{ \boldsymbol{\sigma}}{2i} \right],
\end{eqnarray}
where $P$ represents the path ordering {and} $\mathcal{C}$ is a closed
loop on the {$\alpha$ plane}. An essential observation is that for loops
$\mathcal{C}$ that are invariant under TRS (i.e., loops that are
mapped onto themselves by TRS {up to} orientation),
the sewing condition
(\ref{eq:sewing condition for U2 guage field})
quantizes
the SU(2) Wilson loops, $W[\mathcal{C}]=\pm 1$. For example, for a
straight loop $\mathcal{C}_1$ running from $(-\pi,0)$ to $(\pi,0)$
(Fig.\ \ref{fig:3}),
$
W[\mathcal{C}_1]
=
\mathrm{Pf}\left[\tilde{w}(\pi,0)\right]
\mathrm{Pf}\left[\tilde{w}(0,0)\right]$.

Although being invariant under a
gauge transformation induced by a unitary transformation of the local
frame at $\alpha$, each Wilson loop is not invariant under redefinition
of unit cells. For example, when there are only two orbitals ($t=1,2$)
in each unit cell, a {spin-dependent} redefinition of unit cell
$
c_{r,1,s} \to c_{r,1,s}
$,
$
c_{r,2,s} \to
c_{r+ s p \hat{\mathbf{x}},2,s}
$,
$p\in Z$
flips the sign of $W[\mathcal{C}_1]$ and $W[\mathcal{C}_2]$ when $p$ is
odd, where $\hat{{\bf x}}$ is a unit lattice translation vector in
{$x$ direction}.
In other word,
this transformation
inserts a half unit of the SU(2)
flux along the $\alpha_y$-direction. When there is translation
invariance, the choice of unit cells is arbitrary, whereas if we
introduce a boundary, the choice of unit cell should be consistent with
the location of the boundary. In this sense, the (quantized) value of
each Wilson loop matters when we terminate the system.

We can construct, however, from two parallel Wilson loops
(e.g., $\mathcal{C}_1$ and $\mathcal{C}_2$),
an invariant
which is left unchanged by the half-flux insertion in $\alpha_x$ and
$\alpha_y$ directions, since the effect of the half-flux insertion
cancels. We are thus led to consider a Wilson loop that runs
{counterclockwise} around the boundary of the half of $T^{2}_{\alpha}$,
$T^{2}_{\alpha}/2 := (-\pi,\pi]\times [0,\pi]$, say (Fig.\ \ref{fig:3}):
\begin{align}
W[\partial(T^{2}_{\alpha}/2) ]
&:=
(-1)^{\Delta}
=
\!\!
\prod_{
\substack{
k=(0,0),(\pi,\pi),\\
\,\, (\pi,0),(0,\pi)
}
\! \! }
\mathrm{Pf}\left[\tilde{w}(k)\right].
\end{align}
The ${Z}_2$ number $\Delta$ distinguishes trivial ($\Delta=0$) and
non-trivial ($\Delta=1$) insulators. The Kane and Mele
model\cite{KaneMele} is an explicit example of the latter case.

The connection between the gluing and the gauge pictures can
be established for the non-interacting case
with conserved $S^z$.
In this case, the ground state can be written as
$|\Psi^{\alpha}\rangle
= |\Psi^{\alpha}_\uparrow \rangle \times | \Psi^{\alpha}_\downarrow\rangle$
and the doublet becomes
$v(\alpha)
= ( | \Psi^{\alpha}_\downarrow\rangle,
|\Psi^{\alpha}_\uparrow \rangle )$.
The $Z_2$ invariant obtained in the gluing picture is nothing
but $(-1)^{\mathrm{Ch}_{\uparrow}} = (-1)^{\mathrm{Ch}_{\downarrow}}$.
On the other hand, the quantized SU(2) Wilson loop also gives
$W[\partial(T^{2}_{\alpha}/2)] = (-1)^{\mathrm{Ch}_{\uparrow}}$.
The $Z_2$ invariant is
independent of the choice of sublattice in Eq. (\ref{eq:1})
because annihilating $N$ electrons of spin $s$ results in
the same state of $|\Psi_{-s}^\alpha \rangle$
irrespective of the choice of sublattice.
Since the quantized $Z_2$ number cannot change abruptly as we turn
on interactions or disorders, it is the only topological
invariant in weakly interacting TRS systems.
Thus the $Z_2$ invariant is a natural generalization of the parity
(even/odd) of the Chern number, to the case without the $S^z$
conservation in which case the Chern number cannot be defined.

Finally, a few remarks are in order.
First, if we add some perturbations which
enlarge the size of the unit cell from $m=2$ to $m>2$,
we can still remove half of the total electrons at
the original location to calculate the $Z_2$ invariant.
This allows one to characterize the topological order of systems
which have both interactions and disorder
\cite{Essin07}.
Second, the formalism developed so far can be extended to
three dimensions similarly
\cite{Moore06,Roy06_3Da,Fu06_3Da,Fu06_3Db}.

This work has been supported by the National Science Foundation under
Grant No.\ PHY05-51164 (SL and SR) and NSERC (SL).
We thank
L. Balents,
C. Kane,
Y. B. Kim,
J. Moore,
R. Roy
and
A. Vishwanath for helpful {comments and discussions.}


\begin{thebibliography}{99}
\bibitem{convention}
We use the convention where a quantum operator
(which is `hatted') acts
on a doublet of quantum states
from the left-hand side,
whereas a c-number matrix acts
from the right-hand side,
e.g., $( \hat O v M )_n =
\sum_m \hat O |v_m \rangle M_{mn}$.
\end{thebibliography}
\end{document}